# Temperature dependence of the dielectric permittivity of $CaF_2$, $BaF_2$ and $Al_2O_3$ : Application to the prediction of a temperature dependent van der Waals surface interaction exerted onto a neighbouring Cs ($8P_{3/2}$) atom

**Thierry Passerat de Silans[1], Isabelle Maurin[1], Pedro Chaves de Souza Segundo[1], Solomon Saltiel[1,3], Marie-Pascale Gorza[1], Martial Ducloy[1], Daniel Bloch[1,*], Domingos de Sousa Meneses[2,4], Patrick Echegut[2]**

[1] *Laboratoire de Physique des Lasers, UMR7538 du CNRS et de l'Université Paris13, 99 av. JB Clément, Villetaneuse, France*
[2] *CNRS, UPR3079 CEMHTI, 1D Avenue de la Recherche Scientifique, 45071 Orléans cedex 2, France*

*Abstract    The temperature behaviour in the range 22°C to 500 °C of the dielectric permittivity in the infrared range is investigated for $CaF_2$, $BaF_2$ and $Al_2O_3$ through reflectivity measurements. The dielectric permittivity is retrieved by fitting reflectivity spectra with a model taking into account multiphonon contributions. The results extrapolated from the measurements are applied to predict a temperature-dependent atom-surface van der Waals interaction. We specifically consider as the atom of interest Cs ($8P_{3/2}$), the most relevant virtual couplings of which, fall in the range of thermal radiation and are located in the vicinity of the reststrahlen band of fluoride materials.*

[*] daniel.bloch@univ-paris13.fr
[3] *Also at Physics Department, Sofia University, 5 J. Bourchier Boulevard, 1164 Sofia, Bulgaria*
[4] *Also at Université d'Orléans, Polytech, Avenue du Parc Floral, BP 6749, 45067 Orléans cedex 2, France*

1. **Introduction**

Materials that are currently used as optical windows have usually been studied in depth, notably to determine their refractive indices in their transparency regions, and the shift of the edges of transparency region with temperature. In the absorption band, optical investigations of their properties have been commonly performed to get information on their phonon bands, not requiring the same level of accuracy than in common optical applications. However, the optical features of the surface modes, although derived from the knowledge of the bulk material, require information on these optical properties within the absorption band. A sensitive tool of these surface properties – benefiting of the high-accuracy of atomic physics - is provided by the spectroscopy of an atom in the close vicinity with a surface, through the van der Waals type long range surface interaction (see [1] and refs. therein). We have recently addressed [2] the problem of the accuracy of the data existing in the literature for this kind of specific experiments and in particular, we have shown the high sensitivity of these surface responses to a minor inaccuracy in the bulk properties. This explains that to investigate properties connected to the thermal emissivity of a surface [3 - 6], an accurate knowledge of the temperature dependence of the optical properties of the material is needed in the thermal infrared range.

In this work, we investigate the temperature dependence of the infrared properties of fluoride crystals $CaF_2$, and $BaF_2$. The transparency of these materials extends from UV to mid-IR ranges, and their reststrahlen band, corresponding to a strong absorption, lies in the far infrared range. Additionally, we perform also measurements on sapphire ($Al_2O_3$, perpendicular *c*-axis), a well-known optical window material that is considered as a reference for the optical properties, and for the planned experiments. To our best knowledge, the literature results concerning $CaF_2$ are limited to ambient temperature [7, 8, 9] and 100 K [8] range. For $BaF_2$ [7, 10, 11], reflectance data were given for the following temperatures : 295, 373, 573 and 773 K. Large discrepancies between published results and their low resolution justify the realization of new measurements.

For $Al_2O_3$, systematic studies as a function of temperature had already been conducted [12, 13]. The IR behavior is usually described with the help of several (up to 4) phonon modes. However, the experimental findings vary, probably due to variations in the quality of the polishing of the sample, the degree of the impurities and the precision of the measurements.

Our aim in this work is to determine from the knowledge of reflectivity spectra the complex surface response:

$$S(\omega) = \frac{\varepsilon(\omega) - 1}{\varepsilon(\omega) + 1} \qquad (1)$$

where $\varepsilon(\omega)$ represents the dielectric function. This research was triggered by an experimental search of the temperature dependence of the atom-surface interaction [4, 5], in which the fundamental effect of vacuum field temperature is susceptible to combine with the temperature dependence of the dense material. As this experiment in progress is performed onto the $8P_{3/2}$ level of Cs [6,14], for which virtual absorptive couplings to Cs ($7D_{3/2}$) and to Cs ($7D_{5/2}$) lie respectively at ~ 256.4 cm$^{-1}$ (~ 39 μm) and at ~ 277.8 cm$^{-1}$ (~ 36 μm), we apply our temperature-dependent evaluations of S (ω) to the relevant predictions for this specific problem of the influence of the thermal emissivity of materials onto the atom-surface interaction. Note that in the principle, extrapolation of surface properties from information from the bulk assumes that the surface is of the same nature as the bulk: let us note that here, surface changes (impurities, crystal reconstruction) occur on a thickness that is truly negligible relatively to the thermal infrared wavelengths that we consider. In addition, deriving the bulk properties from a reflectivity measurement implicitly assumes this similar nature of the material in the bulk and at the interface.

The paper is presented in the following way. In section II, we briefly describe the experimental results with reflectivity spectra as a function of temperature. In section III, we present the theoretical model [15] of dielectric permittivity that we use to fit the spectra. In section IV, we discuss additional criteria to make individual fittings of a given experimental curve consistent with the expected temperature behaviour, and we provide our estimates of the dielectric permittivity spectra, and of the spectrum of the surface response as a function of temperature. In section V, we present the results of the calculation of the van der Waals attraction coefficient for Cs ($8P_{3/2}$) in front of surfaces of CaF$_2$,

BaF$_2$ and Al$_2$O$_3$ as a function of temperature. By comparing the predictions resulting from various models of permittivity, not limited to our findings, we illustrate why a sensitive knowledge of the temperature-dependent permittivity is required.

2.   **Experimental details**

The samples were optically polished on both sides, with a diameter ~ 8 mm and a 0.8 mm thickness for CaF$_2$ and BaF$_2$, and a 1 mm thickness for Al$_2$O$_3$ (with *c*-axis perpendicular to the window). The reflectivity spectra of CaF$_2$, BaF$_2$ and Al$_2$O$_3$ in infrared range were acquired with a Bruker Infrared Fourier Spectrometer (IFS) 113v interferometer for various temperatures ranging between room temperature and 500 °C by steps of 100 °C. The heating was provided by heating a ceramic plate mounted in contact with the rear facet of the sample. The experimental resolution spectrum was ~ 4 cm$^{-1}$ and the accuracy on the measured reflectivity was ~ 2-3 %.

To extract optical information on the bulk properties of the material, one should note that as long as the sample is absorbing in the frequency range that is analyzed, the reflected intensity only originates in the reflectivity from the input window. Conversely, when the absorption along the sample remains weak, the reflected light should include the reflection from the second window. However, the obtained reflected intensity also combines reflectivity of the sample with the thermal emission of the ceramic hotplate.

The measured reflectivity spectra are reported in figures 1-3. For CaF$_2$, at low temperatures, the spectra exhibit small structures inside the reststrahlen band, which disappear when the temperature increases (see figure 1). These small oscillations, although not consistently mentioned in the literature because of a lower resolution and sensitivity, do not result from interferences nor from an artifact. They were already observed in [6, 8]. Such oscillations are clearly the multiphonon process signature, as explained in [8]. For BaF$_2$ (see figure 2), we also observe the multiphonon contribution within the reststrahlen band. For Al$_2$O$_3$ (see figure 3), the reflectance spectra are more complex and present smaller oscillations also due to multiphonon processes in the reststrahlen region. As a general trend for all these spectra, the sharp edges of the reflectivity get smoother with increasing temperatures.

One also notes on the spectra some apparent oscillations on the wings (50-100 cm$^{-1}$, and 500 cm$^{-1}$ for CaF$_2$ and BaF$_2$ or 1000 cm$^{-1}$ for Al$_2$O$_3$). They must be attributed to noise: the reflectivity being weak and the detector sensibility low, the signal to noise ratio becomes weak.

### 3. Dielectric constant : theoretical background

In the frame of classical dielectric models [13, 16], the dielectric function $\varepsilon$ can be described by:

$$\varepsilon(\omega) = \varepsilon_\infty + \sum_j \frac{S_j \Omega_j^2}{\Omega_j^2 - \omega^2 - i\gamma_j \omega} \qquad (2)$$

where $\varepsilon_\infty$, $\Omega_j$, $S_j$ and $\gamma_j$ are respectively the high-frequency value of the dielectric constant, and the transverse optical wave number, the dielectric strength and the damping of the j$^{th}$ phonon. Such a description does not take into account for the lattice anharmonicity that makes crystal vibrations interacting with the phonon bath in a more complex way: in particular, the damping should not simply be a constant $\gamma_j$, but should be highly frequency-dependent, which is due to multi-phonon interactions (see [15] and refs. therein). The effect of this multi-phonon interaction was addressed in several models [10, 15]. It can be notably taken into account in an advanced model [15] where $\gamma_j$ is replaced by a self-energy function $P_j(\omega)$, consisting of a sequence of peaks that are well reproduced by "extended" Gaussian functions (*i.e.* with Kramers-Kronig counterparts) :

$$\varepsilon(\omega) = \varepsilon_\infty + \sum_j \frac{S_j \Omega_j^2}{\Omega_j^2 - \omega^2 - 2\Omega_j P_j(\omega)} \qquad (3)$$

with: $P_j(\omega) = \sum_n \tilde{g}_{n,j}(\omega)$

$$\tilde{g}_{n,j}(\omega) = \frac{2A_{n,j}}{\sqrt{\pi}} \left[ D\left( \frac{2\sqrt{Ln(2)}(\omega + \omega_{n,j}^0)}{\gamma_{n,j}} \right) - D\left( \frac{2\sqrt{Ln(2)}(\omega - \omega_{n,j}^0)}{\gamma_{n,j}} \right) \right]$$

$$+ iA_{n,j} \exp\left( -\frac{4Ln(2)(\omega - \omega_{n,j}^0)^2}{(\gamma_{n,j})^2} \right) - iA_{n,j} \exp\left( -\frac{4Ln(2)(\omega + \omega_{n,j}^0)^2}{(\gamma_{n,j})^2} \right)$$

(4)

and where $A_{n,j}$, $\omega_{n,j}^0$ and $\gamma_{n,j}$ are respectively the amplitude, the location and the damping parameters of Gaussians, and D(x) the Dawson integral which is given by:

$$D(x) = \exp(-x^2) \int_0^x \exp(t^2) dt \qquad (5)$$

Within such a frame, that considerably increases the number of free parameters, a major problem is to identify the right number of parameters, allowing getting an accurate fitting of the experimental spectra, while keeping the number of parameters at a minimum [15]. The objective pursued in the next section is to extract a unique spectrum of permittivity from the reflectivity data, while finding criteria of consistency relatively to the temperature dependence.

**4.     Fitting results and discussion**

All the numerical fitting spectra have been obtained with *Focus,* a curve fitting program developed by one of us (D. d. S. M.) that is in free access on the web [17].

In figures 1-3, in which the experimental spectra are reproduced, we have also plotted our best fits obtained on the basis of the semi-quantum model described in section III. The parameters used for these fits are applied to determine the spectral behaviour of the dielectric permittivity $\varepsilon(\omega)$ (figures 4-6). These parameters are reported in Tables I – III for $CaF_2$, $BaF_2$ and $Al_2O_3$ respectively, and the complex spectra of self-energy are shown in figures 7-9.

For consistency of the fittings with other known criterions, we have selected fittings for which the extrapolation at zero frequency of the dielectric constant $\varepsilon_0$ is in agreement with the one known in literature at ambient temperature (*i.e.* 6.63 for $CaF_2$ [9], 6.7 – 7.4 for $BaF_2$ [11] and 9.395 for $Al_2O_3$ ordinary ray [18]). We also have tried to impose as much as possible that parameters appearing in the dielectric constants and in the self-energy (see Tables I–III and figures 4 – 9) evolve linearly with temperature. Such an *a priori* behaviour can be justified (see *e.g.* [19]). Note that in the absence of such a requirement, that applies to a set of spectra recorded at various temperatures, a single reflectivity spectrum can be equivalently fitted by several systems of parameters, at the expense of the reliability of the extrapolated spectra of permittivity (figures 4-6) or of the surface responses (figures 10-12).

For $CaF_2$ (fig 1), and as predicted by group theory [20], infrared activity only concerns phonon modes with the T1u symmetry which is triply degenerate. So to fit the infrared spectra and their temperature dependence, a dielectric function model involving a single anharmonic phonon term was used. A rather complex self-energy function was also necessary to reproduce the optical response due to the multiphonon processes. *i.e.* the numerous small structures appearing within the reflection band. In spite of some apparently erratic variations of the individual parameters with temperature (Table I), these parameters are needed to take care of the evolution of the multiphonon contributions observed around 400 cm$^{-1}$. Figure 10 shows that the temperature behavior of the overall shape of the self-energy function, is consistent with the tendency that is expected. The peak amplitude of the corresponding dielectric constant (figure 4) decreases by 82 % when the temperature increases from 22 °C to 500 °C. Simultaneously, the resonance corresponding to the maximum of Im[$\varepsilon(\omega)$] shifts from 260 cm$^{-1}$ to 240 cm$^{-1}$.

For BaF$_2$ (figure 2), the fits also involve a single anharmonic phonon term and a self-energy structure to take care of the multiphonon processes. The evolution of the fitting parameters is here clearly monotonous with temperature. The amplitude of the corresponding dielectric constant (figure 5) decreases by 16 % when the temperature increases from 22 °C to 500 °C and the resonance shifts from 185 cm$^{-1}$ to 171 cm$^{-1}$.

For Al$_2$O$_3$ (see figure 3), the fit involves four phonons terms, all with their own self-energy functions, but only a single resonance [called S$_3$ in Table III] exhibits a complex self-energy structure. The amplitude of the corresponding dielectric constant for the main peak decreases by 4.5 % when the temperature increases from 22 °C to 500 °C and the resonance shifts from 443 cm$^{-1}$ to 432 cm$^{-1}$ (figure 6).

## 5. Predictions for the C$_3$(T) van der Waals coefficient for the interaction between Cs (8P$_{1/2}$) and a fluoride window

The long-range atom-surface van der Waals (vW) interaction is a dipole-dipole interaction between the quantum dipole fluctuations of an atom and its electric image induced in the surface, so that an atom is submitted to a potential V(z) = -C$_3$ z$^{-3}$, with z the atom-surface separation and C$_3$ the coefficient of the van der Waals (vW) interaction (*i.e.* attraction as long as C$_3$ > 0). The C$_3$ coefficient is known (for a review, see [21]) to depend notably upon the virtual transitions connecting the atomic level of interest, and upon the surface resonances, determined from the knowledge of the spectrum of the surface response S(ω) :

$$C_3(|i\rangle) = \sum_j r_{ij}(T) \langle i|D|j\rangle^2 \qquad (6)$$

with |i⟩ the considered atomic state, D the dipole operator, and r$_{ij}$ (T) the dielectric image coefficient. This image coefficient, equal to unity for an ideal electrostatic reflector, depends on the surface response spectrum S(ω) and on the vacuum temperature.

The theoretical temperature dependence of $r_{ij}$ (T) and hence of $C_3$(T) was recently analyzed in [5] (see notably eqs 3.10 and 3.11 of this reference), under the assumption that the temperature of the surrounding vacuum field (*i.e.* roughly, the thermal number of photons) is in equilibrium with the one of the dense material. In this approach, the spectrum of the dielectric permittivity is taken as phenomenologically granted. Clearly, to study the evolution over a limited temperature range in [5], and keeping the assumption that the temperature evolution of the vacuum field is relatively decoupled from the details of the spectral modifications of the dense material, the spectrum of dielectric permittivity to be considered should be the one at the relevant temperature.

Below, we apply the results obtained in the preceding sections, in order to predict the temperature dependence in the sensitive case of Cs ($8P_{3/2}$), and to compare it with alternate predictions resulting from various modelling of the dielectric permittivity. Indeed, the $8P_{3/2}$ level of Cs, of a relatively easy experimental access [6, 14], exhibits major couplings in the thermal infrared range, as recalled in the Table IV, reproduced from [14]. As already noticed, surface resonances of fluoride materials fall also in the thermal infrared range, so that strong variations of the $C_3$ coefficient are expected with temperature. In particular, the surface thermal emissivity is susceptible to resonantly couple to a virtual atomic absorption (the frequency position of the most relevant one, at 277.8 cm$^{-1}$ (~ 36 μm), is indicated in figures 10-12), opening up the possibility of a vW repulsion above a certain temperature.

In figures 13-15, we have plotted the predicted $C_3$ values of Cs ($8P_{3/2}$) as a function of temperature. For purposes of comparison, these values are derived from the formulas obtained in [5] (eqs. 3.10 and 3.11) applied with various modellings of permittivity or the dielectric window. Note that because in the eqs. 3.10 and 3.11 of ref. 5, an analytical expansion of the permittivity is required for the non resonant term,– *i.e.* the permittivity must be defined for an imaginary frequency-, we have actually combined, for the model derived from our measurements, the actual resonant contribution, and a non resonant contribution, gently evolving with temperature, estimated from an elementary classical model for the material : indeed, it is only for the knowledge of the resonant term that an accurate estimate of the permittivity is needed. This comparison includes : (i) the temperature dependence solely resulting from the temperature dependence of the vacuum, the dielectric medium properties being those found (in our experiments) at room temperature; (ii) the discrete evaluations of the $C_3$ values resulting from our measurements of reflectivity, performed for a discrete set of temperatures; (iii) and temperature-linearized models. About this last type of modelling, it is justified in current models of the literature to assume that all temperature dependence of optical parameters of given material can be linearized [19], once the optical properties of the material were determined for two temperatures. This approach is based on an extended form of the classical oscillator model (*cf.* eq. 2) assuming:

$$\varepsilon(\omega,T) = \varepsilon_\infty + \sum_j \frac{S_j(T)\Omega_j^2(T)}{\Omega^2_j(T) - \omega^2 - i\gamma_j(T)\omega} \quad (7)$$

with the temperature dependence of the transverse optical wave number $\Omega_j$, of the dielectric strengths, $S_j$, and of the damping of the $j^{th}$ phonon, linearized in the following form :

$$\Omega_j(T) = \Omega_j(T_0) + a_j[T - T_0]$$

$$S_j(T) = S_j(T_0) + b_j[T - T_0]$$

$$\frac{\gamma_j}{\Omega_j}(T) = \frac{\gamma_j}{\Omega_j}(T_0) + c_j[T - T_0] \quad (8)$$

with $T_0$ a reference temperature, and $a_j$, $b_j$, and $c_j$ being constant coefficients.

In this modelling, the high-frequency limiting value of the dielectric constant, $\varepsilon_\infty$ also exhibits a linear temperature-dependence :

$$\varepsilon_\infty(T) = \varepsilon_\infty(T_0) + e[T - T_0] \quad (9)$$

where $e$ is a constant coefficient.

We extrapolate coefficients for linearized temperature-dependence from the results of Denham *et al* [8], who measured fluoride crystals dielectric permittivity at two cold temperatures (100 K, and ~300 K) and we find (taking $T_0$ = 300 K):

for $CaF_2$ : $a_1$ = - 0.03 cm$^{-1}$ K$^{-1}$; $a_2$ = -0.03 cm$^{-1}$ K$^{-1}$; $b_1$=0.00063 K$^{-1}$; $b_2$=0 K$^{-1}$; $c_1$=0.0001 K$^{-1}$; $c_2$ = 0.00025 K$^{-1}$; e = -0.00003 K$^{-1}$.

and for $BaF_2$ : $a_1$ =-0.02 cm$^{-1}$ K$^{-1}$; $a_2$=-0.035 cm$^{-1}$ K$^{-1}$; $b_1$=0.0004 K$^{-1}$; $b_2$=0.00013 K$^{-1}$; $c_1$=0.0002 K$^{-1}$; $c_2$ = 0.0005 K$^{-1}$; e = -0.000045 K$^{-1}$.

Note that in the modelling of [8] –as in [7]- two phonons are actually obtained, the second phonon with a contribution ten times smaller than the first one, because the multiphonon processes through a self-energy are not considered. Although the experimental results at room temperature in [8] slightly differ from those obtained both in [7] and in our experiments, it can be reasonably assumed that the systematic uncertainties or errors in [8] affect only marginally the extrapolated temperature dependence, so that this linearized temperature behaviour can be applied without inconsistency to measurements of [7] or [19].

As shown in figure 10, for $CaF_2$, the atomic transition ($8P_{3/2}$ - $7D_{3/2}$) on which the most sensitive temperature effects are expected is located on a wing of the complex response function. This explains that in figure 13, the temperature behavior of $C_3$ is relatively insensitive to the detailed modelling of the $CaF_2$ response. However, one notes that at the highest temperatures for which we have studied $CaF_2$, we are not able to predict whether vW interaction turns to be repulsive, while this would be a major practical conclusion to be derived in view of the current experiments for $CaF_2$.

For $BaF_2$, the atomic transition ($8P_{3/2}$ -$7D_{3/2}$) falls in closer resonance with the surface response (figure 11), so that the effects with temperature on the van der Waals coefficients should be very important. Indeed, predictions for the temperature dependence of $C_3$ radically differ according to the modeling of the $BaF_2$ response (figure 14). It is in particular clear that the basic model of temperature effects for the vW interaction, although able to predict the fundamental feature of a strong temperature dependence connected to the resonant coupling between the atomic absorption and the thermal emissivity, is far from being sufficient as long as quantitative data are expected. In the refined models for which the temperature dependence of the material properties is also considered, the temperature behaviour for $C_3$ exhibits strong nonlinear variations: this is because the general effect of broadening of the $BaF_2$ surface resonance with increasing temperature is susceptible to be enhanced, or counteracted, by the shift of the resonance.

At last, for our "reference"' material $Al_2O_3$, for which the atomic transitions falling in the thermal infrared ($8P_{3/2}$ - $7D_{3/2}$) are out of resonance of the complex response function S (figure 12), the predicted variation of the $C_3$ coefficient remains marginal (figure 15) in our considered temperature range. The real and imaginary parts at this atomic frequency do not change when the temperature increases.

## 6. Conclusion

Although fluoride materials are usually considered to be simple and well-known materials, it appears that a precise and consistent fitting of reproducible data require an elaborate procedure. These details, often of a minor importance in most of the applications, can actually lead to dramatic effects when the sharp selectivity of Atomic Physics resonance are considered. Our precise study of temperature dependence should provide a basis for a refined comparison between the theory, and the results of an experiment in progress.

**Acknowledgements**

The Paris13 team acknowledges partial support by INTAS South-Caucasus Project (grant 06-1000017-9001).

**Table Captions**

Table I: Parameters of the semi-quantum model used to fit the thermal infrared spectra of $CaF_2$.

Table II: Parameters of the semi-quantum model used to fit the thermal infrared spectra of $BaF_2$

Table III: Parameters of the semi-quantum model used to fit the thermal infrared spectra of $Al_2O_3$.

Table IV: Contribution of each virtual transition – with transition wavelengths indicated a minus sign being used for a virtual emission - to the $C_3$ coefficient for Cs ($8P_{3/2}$), assuming the surface is an ideal reflector [14].

**Table I**

| Temperature (°C) | 22 | 100 | 200 | 300 | 400 | 500 |
|---|---|---|---|---|---|---|
| Global parameters | | | | | | |
| $\varepsilon_\infty$ | 2.02 | 2.019 | 2.07 | 2.11 | 2.1 | 2.16 |
| $S_1$ | 4.18 | 4.04 | 3.95 | 3.89 | 3.83 | 3.67 |
| $\Omega_1$ (cm$^{-1}$) | 272.74 | 278.17 | 283.63 | 288.71 | 291.76 | 299.23 |
| Self-Energy | | | | | | |
| $A_{1,1}$ (cm$^{-1}$) | 9.83 | 8.63 | 5.01 | 3.82 | 14.55 | 12.12 |
| $\omega_{1,1}^0$ (cm$^{-1}$) | 54.67 | 54.5 | 54.5 | 54.5 | 54.5 | 54.5 |
| $\gamma_{1,1}$ (cm$^{-1}$) | 260.19 | 245.23 | 214.99 | 189.55 | 167.09 | 179.98 |
| $A_{2,1}$ (cm$^{-1}$) | 8.65 | 10.15 | 4.77 | 7.67 | 10.88 | 9.25 |
| $\omega_{2,1}^0$ (cm$^{-1}$) | 332.74 | 328.92 | 319.06 | 307.51 | 307.47 | 298.43 |
| $\gamma_{2,1}$ (cm$^{-1}$) | 31.34 | 38.81 | 39.26 | 51.86 | 67.18 | 87.66 |
| $A_{3,1}$ (cm$^{-1}$) | 11.145 | 12.598 | 14.44 | 10.27 | 6.66 | 4.07 |
| $\omega_{3,1}^0$ (cm$^{-1}$) | 370.78 | 370.35 | 357.47 | 345.84 | 352.88 | 352.24 |
| $\gamma_{3,1}$ (cm$^{-1}$) | 58.39 | 54.58 | 89.84 | 65.17 | 56.51 | 79.67 |
| $A_{4,1}$ (cm$^{-1}$) | 2.33 | 3.04 | 0.995 | 6.08 | 3.73 | 1.74 |
| $\omega_0^{41}$ (cm$^{-1}$) | 413.56 | 408.84 | 408.99 | 397.19 | 398.97 | 398.97 |
| $\gamma_{4,1}$ (cm$^{-1}$) | 18.61 | 25.91 | 27.16 | 77.05 | 60.76 | 55.25 |
| $A_{5,1}$ (cm$^{-1}$) | 12.95 | 8.22 | 5.81 | 0.97 | 0.31 | 0.129 |
| $\omega_{5,1}^0$ (cm$^{-1}$) | 448.59 | 440.498 | 441.07 | 443.51 | 443.78 | 443.78 |
| $\gamma_{5,1}$ (cm$^{-1}$) | 104.74 | 74.79 | 80.31 | 32.76 | 9.86 | 6.3 |
| $A_{6,1}$ (cm$^{-1}$) | 9.94 | 23.14 | 29.56 | 38.64 | 46.1 | 51.45 |
| $\omega_{6,1}^0$ (cm$^{-1}$) | 640.11 | 619.098 | 625.31 | 618.89 | 620.18 | 621.83 |
| $\gamma_{6,1}$ (cm$^{-1}$) | 483.71 | 436.706 | 492.69 | 537.695 | 602.62 | 690.66 |

**Table II**

| Temperature (°C) | 22 | 100 | 200 | 300 | 400 | 500 |
|---|---|---|---|---|---|---|
| Global parameters | | | | | | |
| $\varepsilon_\infty$ | 2.12 | 2.12 | 2.12 | 2.11 | 2.115 | 2.11 |
| $S_1$ | 4.38 | 4.07 | 3.799 | 3.52 | 3.24 | 2.99 |
| $\Omega_1 (cm^{-1})$ | 194.14 | 202.71 | 209.42 | 217.77 | 226.32 | 235.02 |
| Self-Energy | | | | | | |
| $A_{1,1} (cm^{-1})$ | 5.13 | 3.91 | 2.79 | 2.078 | 1.45 | 0.86 |
| $\omega_{1,1}^0 (cm^{-1})$ | 132.23 | 131.107 | 124.61 | 133.28 | 128.533 | 123.53 |
| $\gamma_{1,1} (cm^{-1})$ | 142.04 | 136.54 | 136.08 | 127.26 | 128.34 | 124.56 |
| $A_{2,1} (cm^{-1})$ | 6.9 | 5.63 | 4.64 | 3.53 | 4.21 | 5.24 |
| $\omega_{2,1}^0 (cm^{-1})$ | 278.48 | 270.11 | 260.89 | 251.44 | 244.89 | 243.97 |
| $\gamma_{2,1} (cm^{-1})$ | 48.2 | 40.48 | 35.17 | 33.87 | 46.78 | 72.15 |
| $A_{3,1} (cm^{-1})$ | 13.84 | 11.23 | 9.908 | 8.09 | 5.42 | 2.67 |
| $\omega_{3,1}^0 (cm^{-1})$ | 331.79 | 326.13 | 318.63 | 307.496 | 295.39 | 295.89 |
| $\gamma_{3,1} (cm^{-1})$ | 46.69 | 52.66 | 65.34 | 80.05 | 64.18 | 47.398 |
| $A_{4,1} (cm^{-1})$ | 14.97 | 28.36 | 37.02 | 48.67 | 57.38 | 64.83 |

| $\omega_0^{41}$ (cm$^{-1}$) | 520.53 | 501.32 | 494.65 | 491.41 | 480.38 | 481.53 |
| --- | --- | --- | --- | --- | --- | --- |
| $\gamma_{4,1}$ (cm$^{-1}$) | 286.4 | 321.43 | 352.42 | 367.86 | 385.52 | 410.35 |

**Table III**

| Temperature(°C) | 22 | 100 | 200 | 300 | 400 | 500 |
|---|---|---|---|---|---|---|
| $\varepsilon_\infty$ | 3.03 | 3.013 | 3.03 | 2.98 | 3.019 | 3.019 |
| $S_1$ | 0.42 | 0.47 | 0.41 | 0.49 | 0.47 | 0.44 |
| $\Omega_1 (cm^{-1})$ | 373.86 | 371 | 370 | 359.67 | 351.66 | 350.6 |
| $A_{1,1} (cm^{-1})$ | 24.41 | 28.45 | 26.06 | 47.79 | 67.54 | 62.47 |
| $\omega_{1,1}^0 (cm^{-1})$ | 199.33 | 204.6 | 219.02 | 219.02 | 208.88 | 215.22 |
| $\gamma_{1,1} (cm^{-1})$ | 248.99 | 241.5 | 229.88 | 203.74 | 200.32 | 196.77 |
| $S_2$ | 2.73 | 2.74 | 2.63 | 2.53 | 2.45 | 2.41 |
| $\Omega_2 (cm^{-1})$ | 439.71 | 436.34 | 435.06 | 432.09 | 429.92 | 427.39 |
| $A_{1,2} (cm^{-1})$ | 5.68 | 7.87 | 7.86 | 8.86 | 9.49 | 10.3 |
| $\omega_{1,2}^0 (cm^{-1})$ | 423.58 | 419.54 | 416 | 413.48 | 409.06 | 403.75 |
| $\gamma_{1,2} (cm^{-1})$ | 38.51 | 45.62 | 59.5 | 65.28 | 77.7 | 93.32 |
| $S_3$ | 2.87 | 2.81 | 2.87 | 2.81 | 2.83 | 2.82 |
| $\Omega_3 (cm^{-1})$ | 580.42 | 582.2 | 582.98 | 578.97 | 581.61 | 582.95 |
| $A_{1,3} (cm^{-1})$ | 1.43 | 5 | 11.48 | 23.67 | 28.36 | 29.86 |
| $\omega_{1,3}^0 (cm^{-1})$ | 371.13 | 371.61 | 307.94 | 354.27 | 362.12 | 347.12 |
| $\gamma_{1,3} (cm^{-1})$ | 190.22 | 200.24 | 237.29 | 192.02 | 198.47 | 206.02 |

| $A_{2,3}$ (cm$^{-1}$) | 10.12 | 9.12 | 7.86 | 8.09 | 5.63 | 4.75 |
|---|---|---|---|---|---|---|
| $\omega_{2,3}^{0}$ (cm$^{-1}$) | 536.83 | 534.45 | 526.44 | 534.32 | 537.77 | 526.56 |
| $\gamma_{2,3}$ (cm$^{-1}$) | 117.73 | 108.42 | 118.29 | 93.84 | 79.03 | 79.42 |

| $A_{3,3}$ (cm$^{-1}$) | 15.56 | 19.17 | 20.42 | 25.5 | 27.59 | 30 |
|---|---|---|---|---|---|---|
| $\omega_{3,3}^{0}$ (cm$^{-1}$) | 914.59 | 912.06 | 878.14 | 839.97 | 815.08 | 805.6 |
| $\gamma_{3,3}$ (cm$^{-1}$) | 333.25 | 414.24 | 418.35 | 369.58 | 366.08 | 362.4 |

| $A_{4,3}$ (cm$^{-1}$) | 6.13 | 6.29 | 8.99 | 9.36 | 15.55 | 17.81 |
|---|---|---|---|---|---|---|
| $\omega_{4,3}^{0}$ (cm$^{-1}$) | 1299.6 | 1353.68 | 1298.6 | 1333.72 | 1301.19 | 1294.23 |
| $\gamma_{4,3}$ (cm$^{-1}$) | 895.54 | 953.88 | 947.96 | 996.79 | 927.75 | 997.607 |
| | | | | | | |
| $S_4$ | 0.15 | 0.15 | 0.138 | 0.15 | 0.17 | 0.175 |
| $\Omega_4$ (cm$^{-1}$) | 638.35 | 636.59 | 633.87 | 628.1 | 618.72 | 610.88 |

| $A_{1,4}$ (cm$^{-1}$) | 6.77 | 6.98 | 7.19 | 8.78 | 15.72 | 22.99 |
|---|---|---|---|---|---|---|
| $\omega_{1,4}^{0}$ (cm$^{-1}$) | 640.36 | 638.2 | 634.9 | 619.7 | 565.2 | 506.43 |
| $\gamma_{1,4}$ (cm$^{-1}$) | 30.03 | 34.07 | 35.79 | 84.38 | 159.506 | 233.39 |

**Table IV**

| Level (i) | λ (μm) | $C_3^i$ (kHz.μm³) |
|---|---|---|
| $5D_{3/2}$ | -0.89 | < 0.01 |
| $5D_{5/2}$ | -0.89 | 0.01 |
| $6D_{3/2}$ | -3.12 | 0.08 |
| $6D_{5/2}$ | -3.16 | 0.71 |
| **$7D_{3/2}$** | **39.05** | **5.32** |
| **$7D_{5/2}$** | **36.09** | **37.78** |
| $8D_{3/2}$ | 4.95 | 0.42 |
| **$8D_{5/2}$** | **4.92** | **3.70** |
| $9D_{3/2}$ | 3.29 | 0.09 |
| $9D_{5/2}$ | 3.28 | 0.80 |
| $10D_{3/2}$ | 2.72 | 0.03 |
| $10D_{5/2}$ | 2.72 | 0.31 |
| $11D_{3/2}$ | 2.44 | 0.02 |
| $11D_{5/2}$ | 2.43 | 0.16 |
| $12D_{3/2}$ | 2.27 | 0.01 |
| $12D_{5/2}$ | 2.27 | 0.09 |
| $6S_{1/2}$ | -0.39 | 0.03 |
| $7S_{1/2}$ | -1.38 | 0.31 |
| **$8S_{1/2}$** | **-6.78** | **12.07** |
| **$9S_{1/2}$** | **8.94** | **11.63** |
| $10S_{1/2}$ | 3.99 | 0.38 |
| $11S_{1/2}$ | 3.00 | 0.09 |
| $12S_{1/2}$ | 2.58 | 0.04 |
| $C_3 = \sum_i C_3^i =$ | | 73.79 kHz.μm³ |

**Figure Captions**

Figure1: Experimental reflectance spectra of $CaF_2$ and fits obtained with semi-quantum dielectric function models for sample temperatures between 22 °C and 500 °C.

Figure 2: Experimental reflectance spectra of $BaF_2$ and fits obtained with semi-quantum dielectric function models for sample temperatures between 22 °C and 500 °C.

Figure 3: Experimental reflectance spectra of $Al_2O_3$ and fits obtained with semi-quantum dielectric function models for sample temperatures between 22 °C and 500 °C.

Figure 4: Dielectric constant of $CaF_2$ (real and imaginary parts) for sample temperatures between 22 °C and 500 °C.

Figure 5: Dielectric constant of $BaF_2$ (real and imaginary parts) for sample temperatures between 22 °C and 500 °C.

Figure 6: Dielectric constant of $Al_2O_3$ (real and imaginary parts) for sample temperatures between 22 °C and 500 °C.

Figure 7: Self-energy $P(\omega)$ (real and imaginary parts) of $CaF_2$ for sample temperatures between 22°C and 500 °C.

Figure 8: Self-energy $P(\omega)$ (real and imaginary parts) of $BaF_2$ for sample temperatures between 22 °C and 500 °C.

Figure 9: Self-energy $P_3(\omega)$ (real and imaginary parts) of $Al_2O_3$ (for $S_3$ resonance only, see table III) for sample temperatures between 22 °C and 500 °C.

Figure 10: Surface response function $S(\omega)$ of $CaF_2$ (real and imaginary parts) for sample temperatures between 22 °C and 500 °C. The virtual atomic absorption of Cs transition $8P_{3/2} \rightarrow 7D_{3/2}$ is indicated by a vertical dashed line at 277.8 cm$^{-1}$.

Figure 11 Surface response function $S(\omega)$ of $BaF_2$ (real and imaginary parts) for sample temperatures between 22 °C and 500 °C. The virtual atomic absorption of Cs transition $8P_{3/2} \rightarrow 7D_{3/2}$ is indicated by a vertical dashed line at 277.8 cm$^{-1}$.

Figure 12: Surface response function $S(\omega)$ of $Al_2O_3$ (real and imaginary parts) for sample temperatures between 22 °C and 500 °C. The virtual atomic absorption of transition Cs $8P_{3/2} \rightarrow 7D_{3/2}$ is indicated by a vertical dashed line at 277.8 cm$^{-1}$.

Figure 13: Predicted $C_3$ vW coefficient for Cs ($8P_{3/2}$) in front of a $CaF_2$ window as a function of temperature : discrete points (a) : $CaF_2$ permittivity according to the fitting of our temperature-dependent experimental results; curve (b) : $CaF_2$ permittivity according to the room-temperature measurements of [7] ; curve (c) : $CaF_2$ permittivity including linear temperature corrections (extrapolated from [8]) to the room-temperature measurements of [7].

Figure 14: Predicted $C_3$ vW coefficient for Cs ($8P_{3/2}$) in front of a $BaF_2$ window as a function of temperature ; discret points (a) : $BaF_2$ permittivity according to the fitting of our temperature-dependent experimental results; curve (b) : $BaF_2$ permittivity according to the room-temperature measurements of [7]; curve (c) : $BaF_2$ permittivity including linear temperature corrections (extrapolated from [8]) to the room-temperature measurements of [7] ; curve (d) : $BaF_2$ permittivity including linear temperature corrections (extrapolated from [8]) to the room-temperature measurements of [8].

Figure 15: Predicted $C_3$ vW coefficient for Cs ($8P_{3/2}$) in front of a $Al_2O_3$ window as a function of temperature: the $Al_2O_3$ permittivity is according to the fitting of our temperature-dependent experimental results.

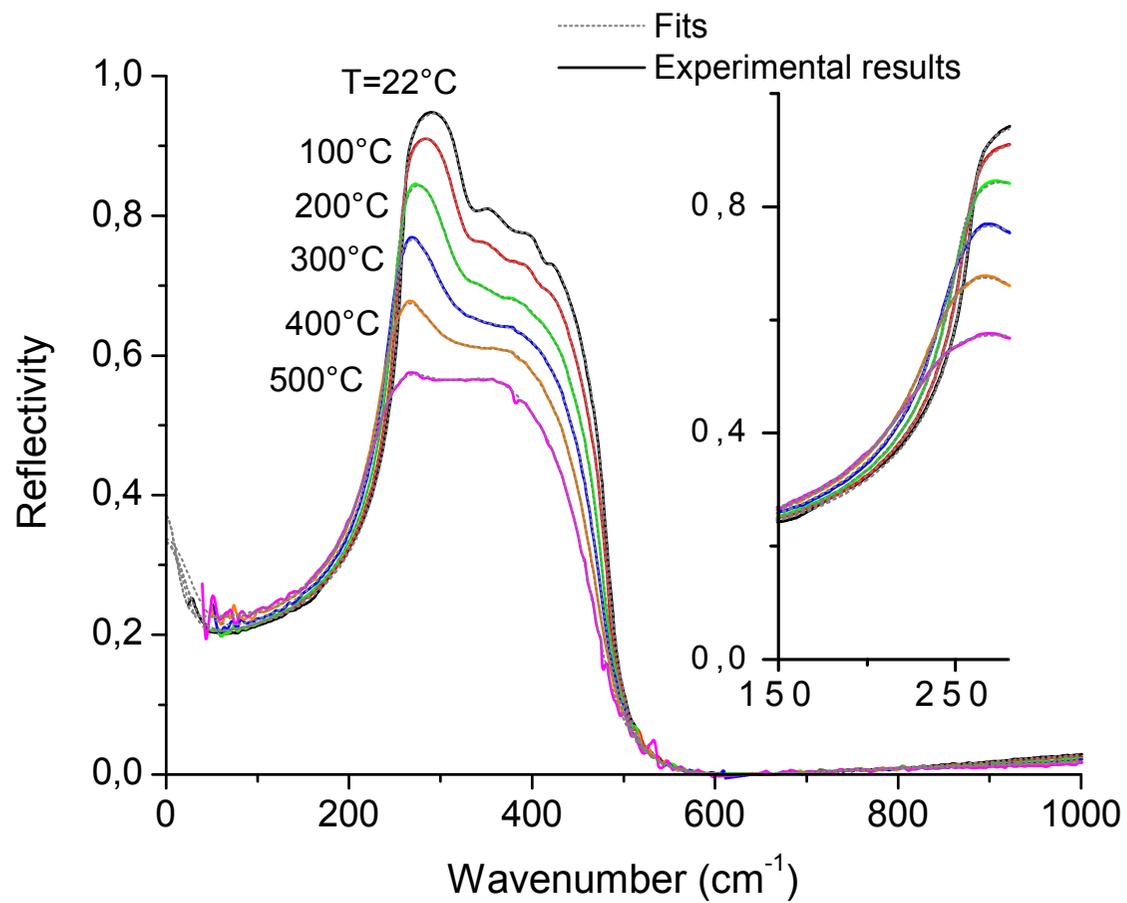

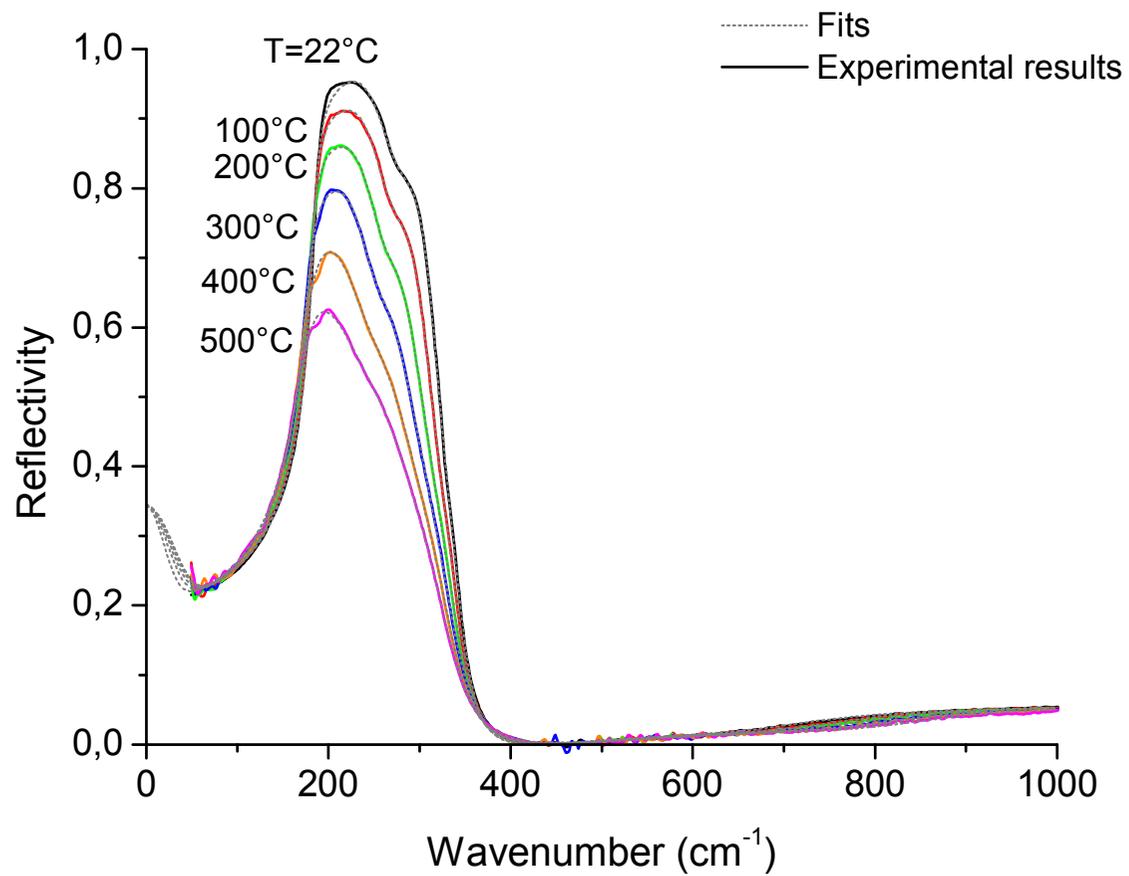

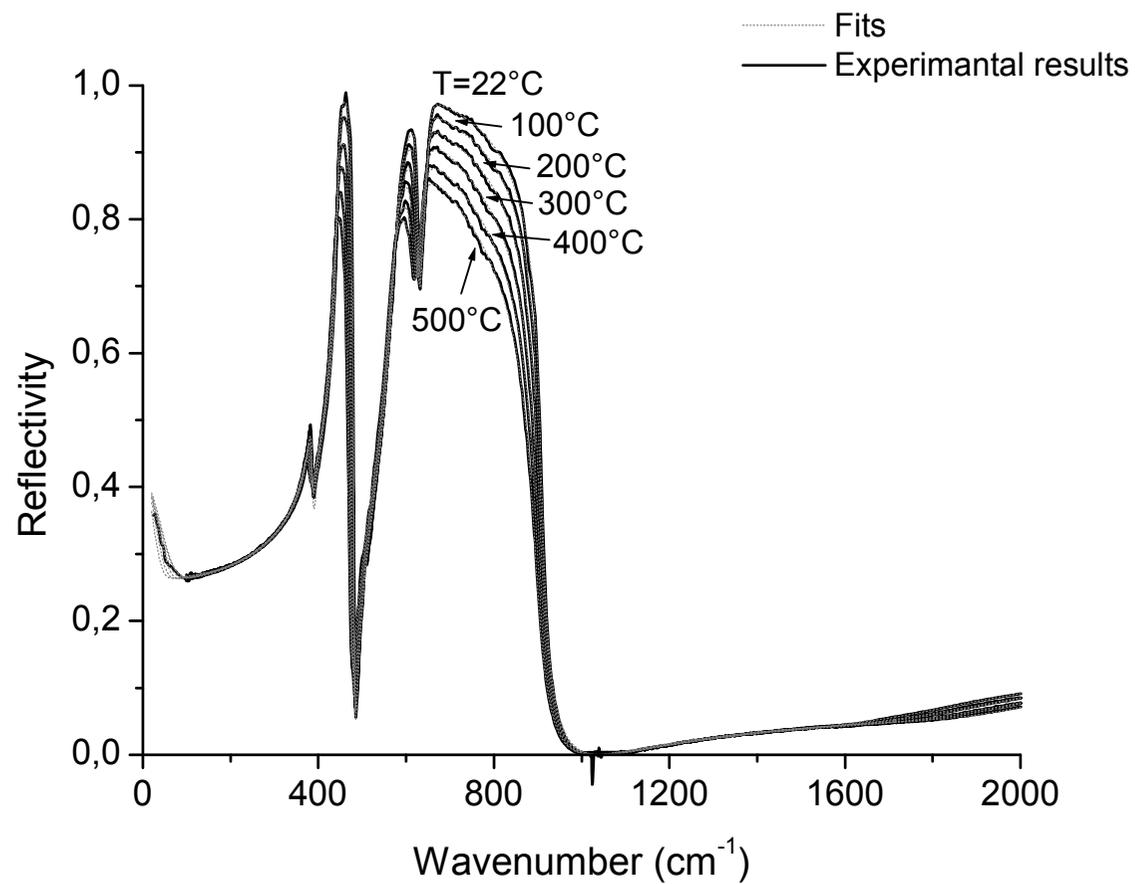

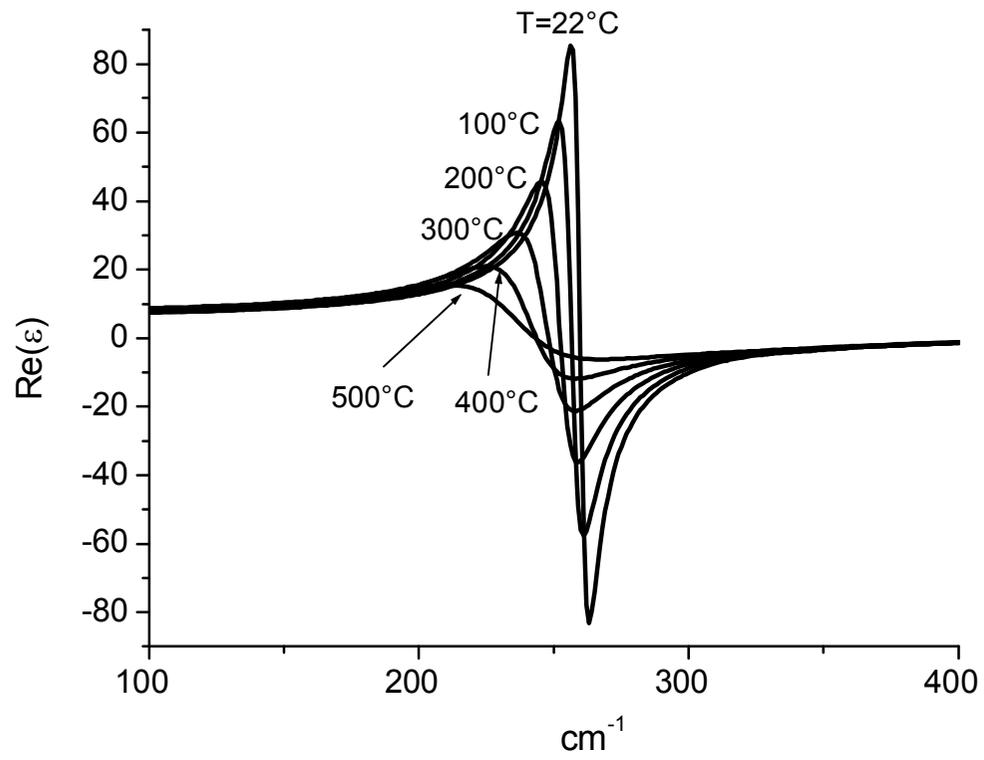
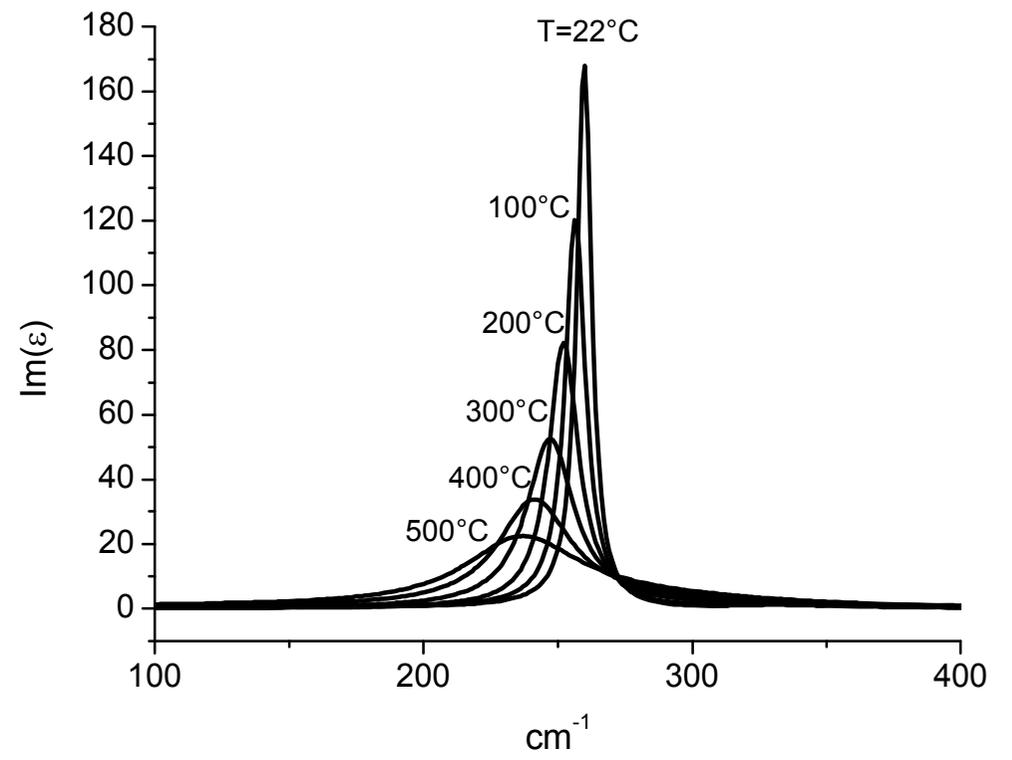

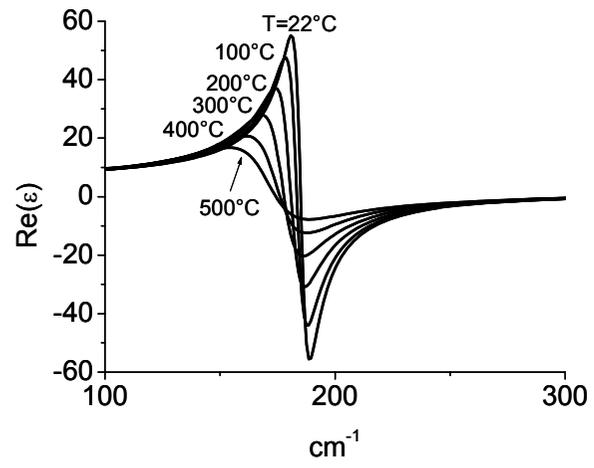 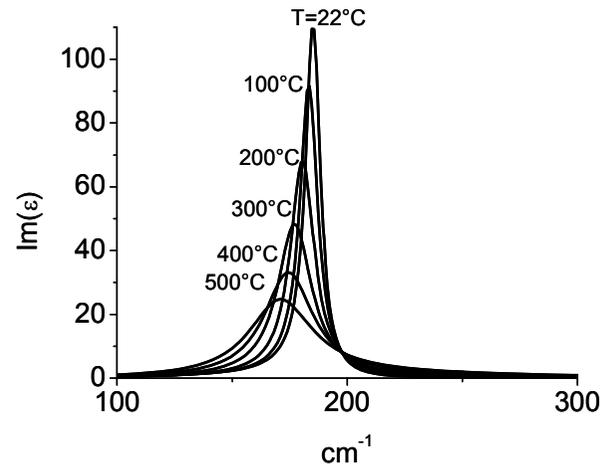

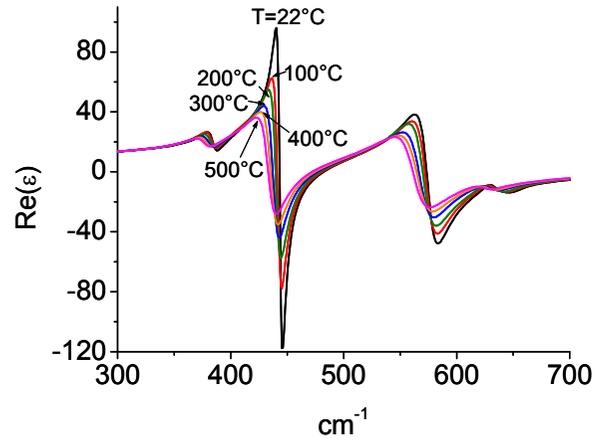 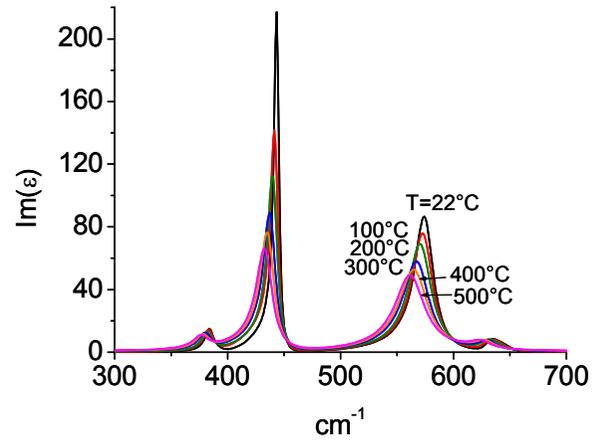

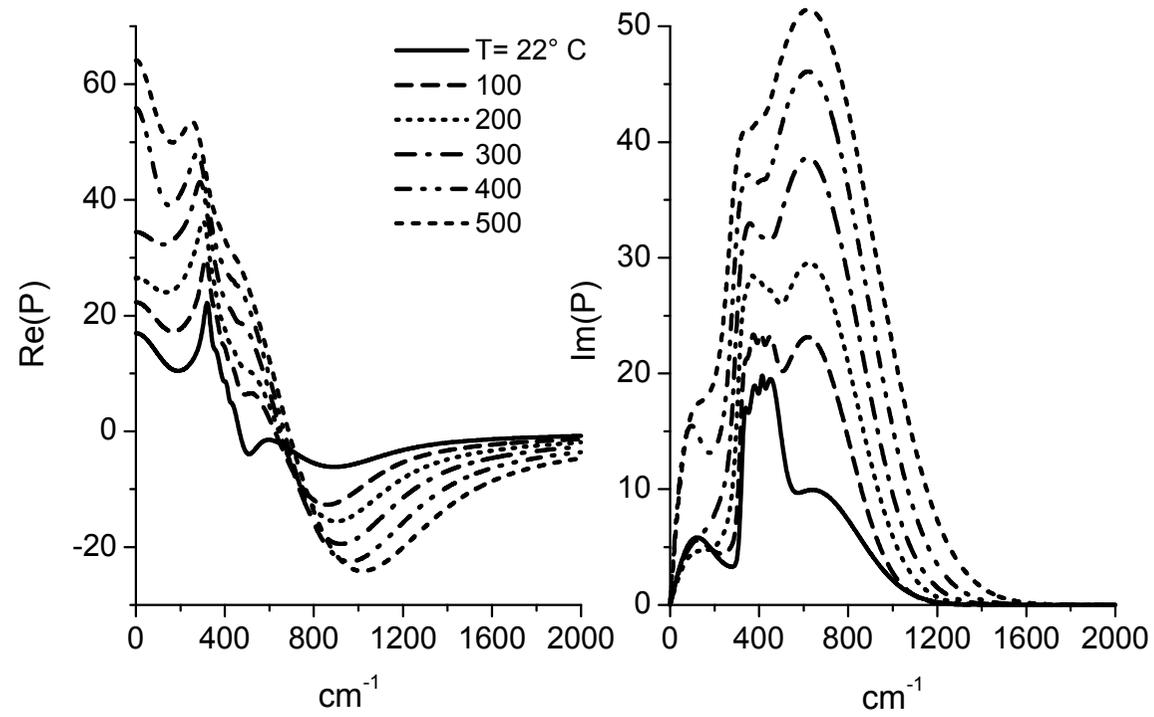

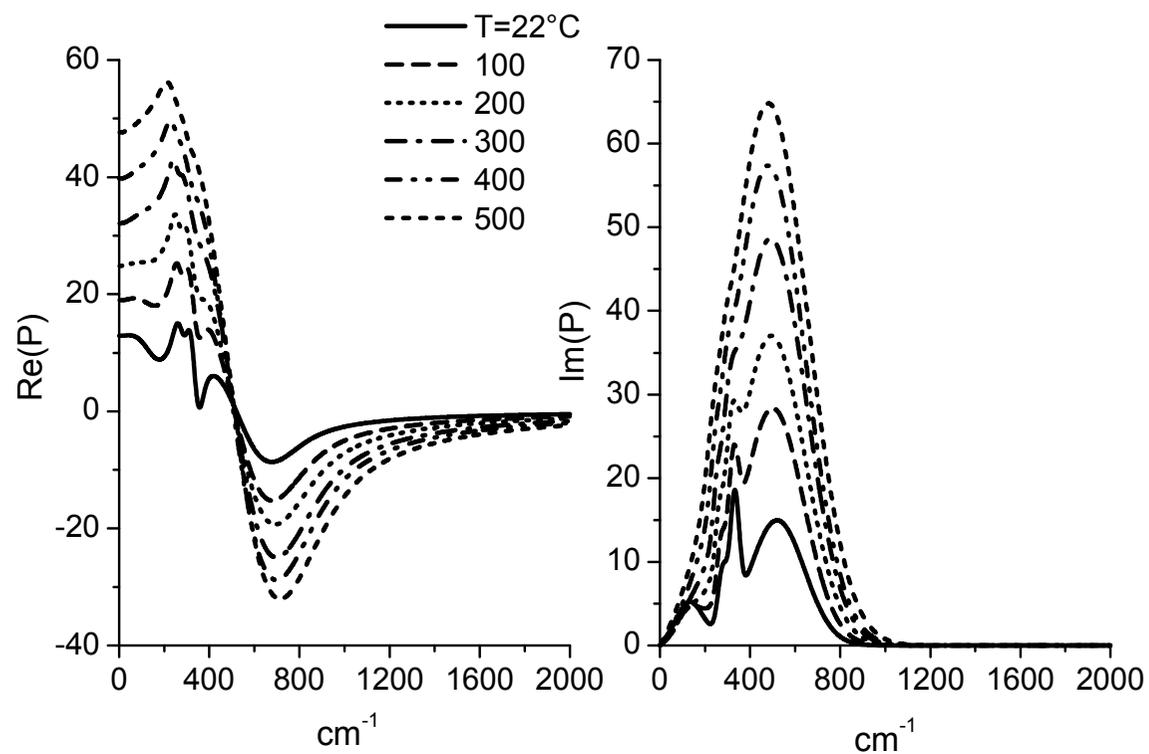

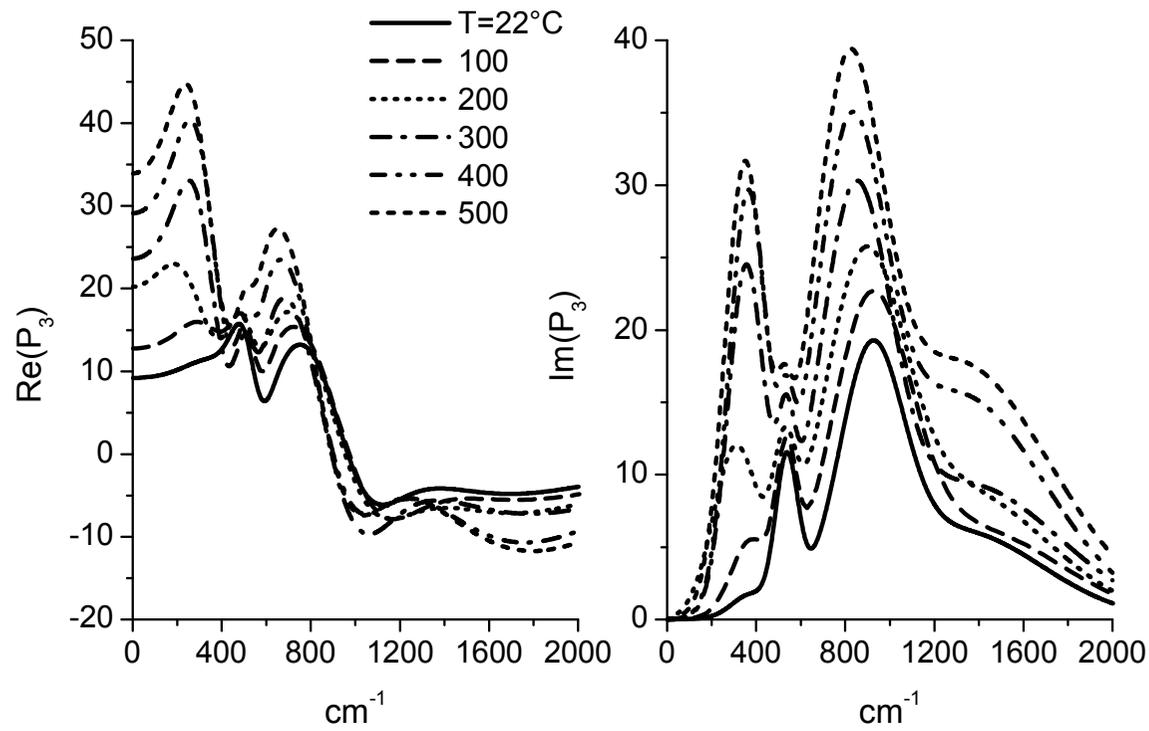

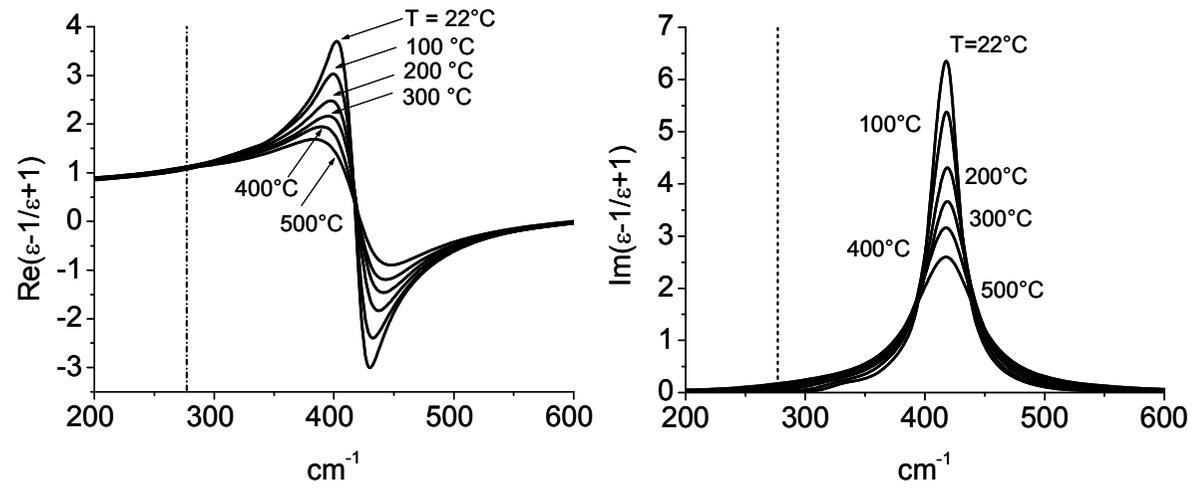

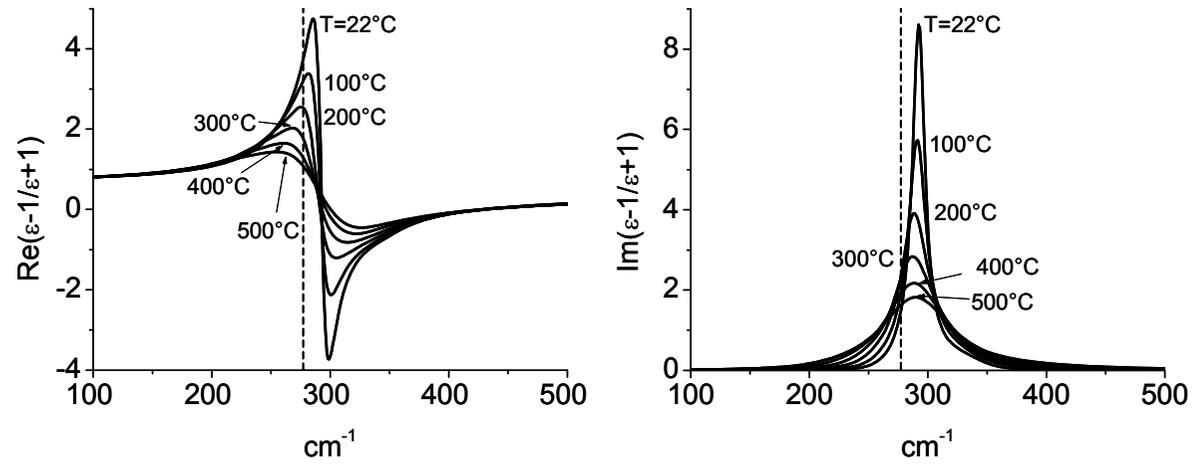

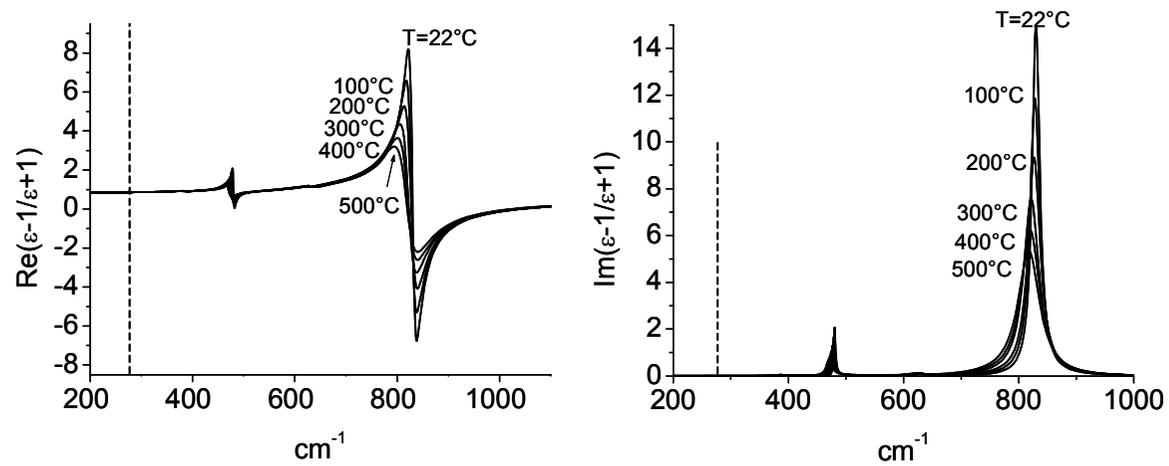

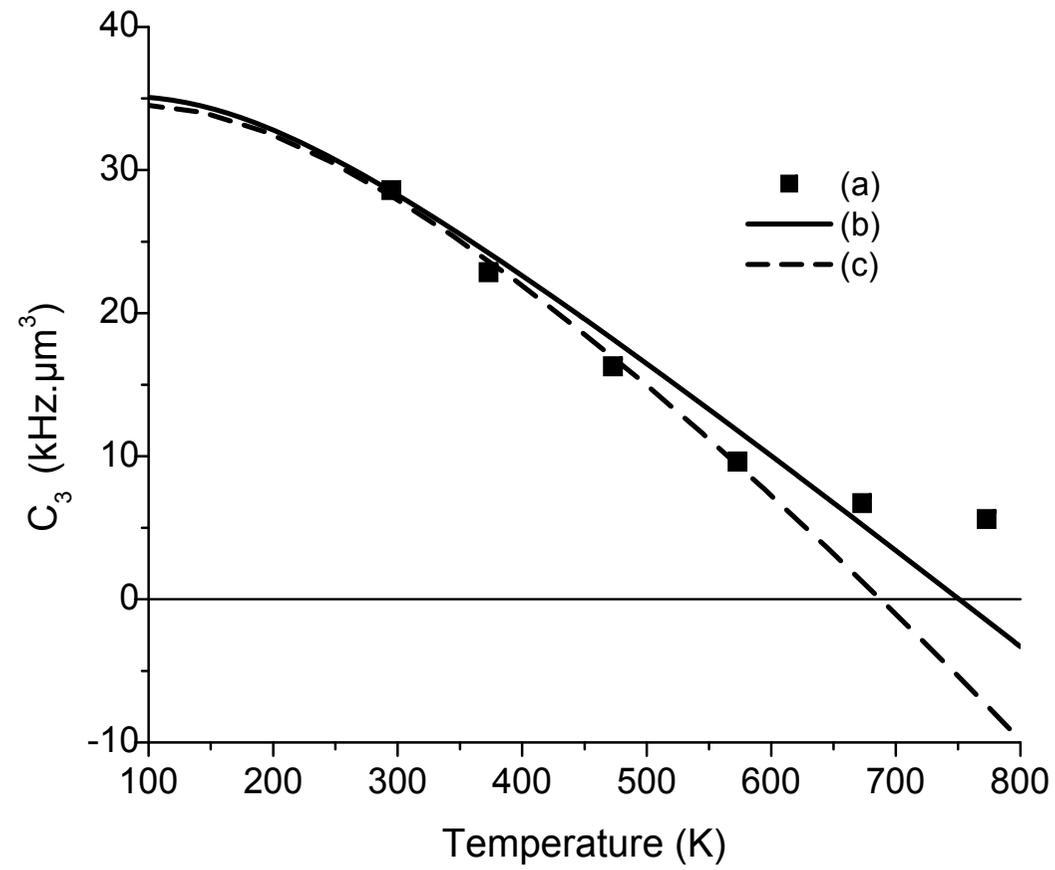

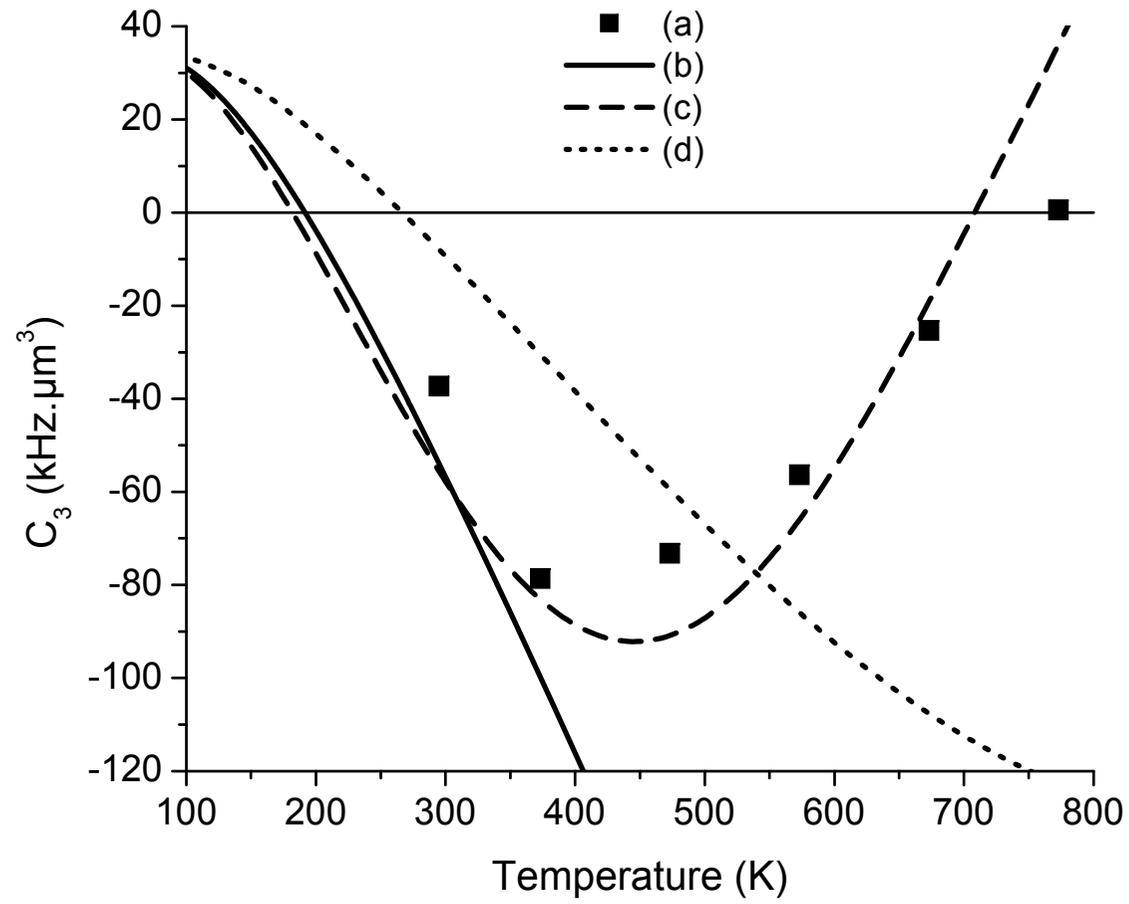

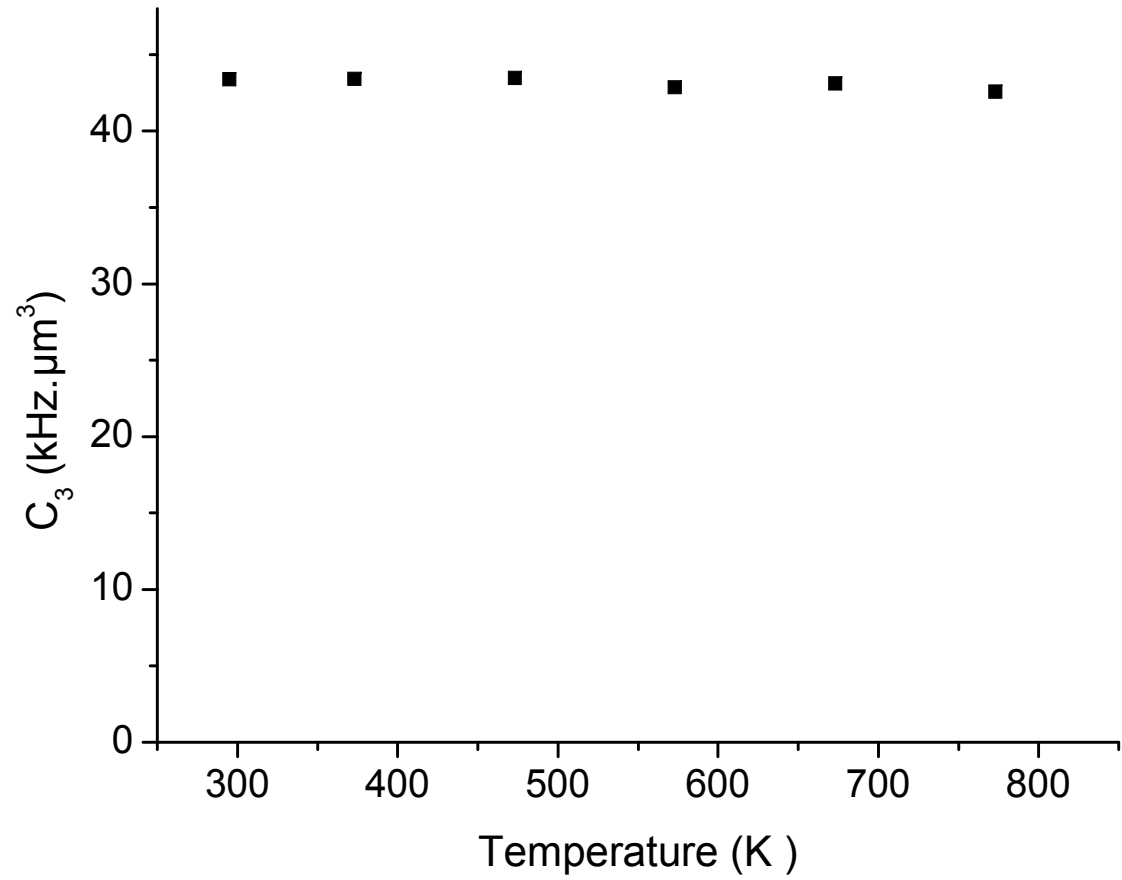